\documentclass[12pt]{article}
\usepackage{a4wide}
\usepackage{amssymb}
\usepackage{amsmath}
\usepackage{graphicx}
\usepackage{xspace}
\usepackage{epsfig}
\usepackage{url}

\newcommand{\ee}{e^+e^-}

\newcommand{\as}{\alpha_s}
\newcommand{\aew}{\alpha_{EW}}

\newcommand{\order}[1]{\mathcal{O}\left(#1\right)}
\newcommand{\cf}{cf.\ }
\newcommand{\ie}{i.e.\ }
\newcommand{\eg}{e.g.\ }

\newcommand{\GeV}{\,\mathrm{GeV}}
\newcommand{\TeV}{\,\mathrm{TeV}}
\newcommand{\fb}{\,\mathrm{fb}}

\newcommand{\bb}{{\bar b}}
\newcommand{\pt}{{p_t}}
\newcommand{\Pt}{{P_t}}
\newcommand{\mb}{{m_b}}

\newcommand{\cO}[1]{\mathcal{O}\left(#1\right)}

\begin{document}

\titlepage
\begin{flushright}
Bicocca-FT-07-5\\
CERN-PH-TH-07-067\\
April 2007\\
\end{flushright}

\vspace*{0.3in}
\begin{center}
{\Large \textbf{\textsf{
      Accurate QCD predictions for heavy-quark jets at the Tevatron and LHC 
}}}\\
\vspace*{0.4in}
A.~Banfi$^{(a)}$,
G.~P.~Salam$^{(b)}$,
and G.~Zanderighi$^{(c)}$ \\
{\small
\vspace*{0.5cm}
$^{(a)}$ {\it Universit\`a degli Studi di Milano-Bicocca\\
and INFN, Sezione di Milano-Bicocca, Italy;} \\
\vskip 2mm
$^{(b)}$ {\it LPTHE,   Universit\'e Pierre et Marie Curie -- Paris 6,
  Universit\'e Denis Diderot -- Paris 7,
  CNRS UMR 7589, 75252 Paris cedex 05, France;}\\
\vskip 2mm}
$^{(c)}$ {\it Theory Division, Physics Department, CERN, 1211 Geneva
  23, Switzerland.
}\\
\vskip 2mm
\end{center}

\vspace{0.5cm}
\begin{abstract}
  Heavy-quark jets are important in many of today's collider studies
  and searches, yet predictions for them are subject to much larger
  uncertainties than for light jets. This is because of strong
  enhancements in higher orders from large logarithms, $\ln
  (\pt/m_Q)$. We propose a new definition of heavy-quark jets, which
  is free of final-state logarithms to all orders and such
  that all initial-state collinear logarithms can be resummed into the
  heavy-quark parton distributions.
  Heavy-jet spectra can then be calculated in the massless
  approximation, which is simpler than a massive calculation and
  reduces the theoretical uncertainties by a factor of three.
  This provides the first ever accurate predictions for inclusive $b$-
  and $c$-jets, and the latter have significant discriminatory power for
  the intrinsic charm content of the proton.
  The techniques introduced here could be used to obtain heavy-flavour
  jet results from existing massless next-to-leading order
  calculations for a wide range of processes.
  We also discuss the experimental applicability of our flavoured jet
  definition.
\end{abstract}
\vspace{0.5cm}

\section{Introduction}
\label{sec:intro}
Studies of heavy-quark jets, \ie charm and bottom jets, are important
for a range of reasons. They are of intrinsic interest because charm
and bottom are the flavours for which there exists the most direct
correspondence between parton level production and the observed hadron
level.
They have the potential to provide information on the $c$- and
$b$-quark parton distribution function (PDF), which are the only
components of proton structure that are thought to be generated entirely
perturbatively from the DGLAP evolution of the other flavours.
Furthermore, $b$-jets enter in many collider searches, notably because
they are produced in the decays of various heavy particles, \eg top
quarks, the Higgs boson (if light) and numerous particles appearing in
proposed extensions of the Standard Model (SM)~\cite{TDR}.

Within the SM a range of production channels exist for heavy-quark
jets, \eg pure QCD production or in association with heavy bosons ($W,
Z, H\ldots$), see \eg \cite{bplusheavy}.
The simplest and most fundamental measurement of heavy-quark jet
production is the inclusive heavy-quark jet spectrum, which is
dominated by pure QCD contributions.
Predictions for this sort of quantity have always been obtained using
calculations in which the $c$ or $b$ quark has been explicitly taken
to be massive while all other lighter masses are neglected.

\begin{figure}[t]
    \centering
    \includegraphics[width=0.7\textwidth]{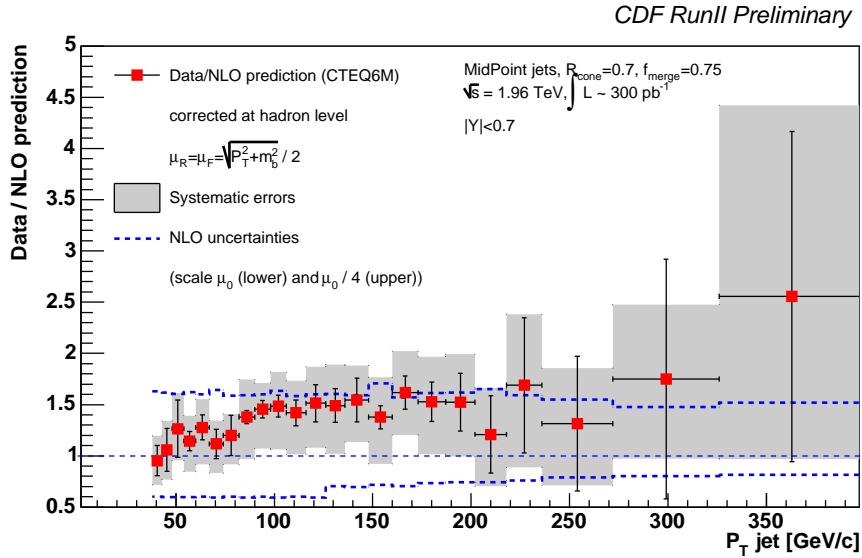}
  \caption{Ratio of the measured inclusive $b$-jet spectrum to NLO prediction.
    The measurement is performed for jets with transverse momentum
    38~GeV $< P_{T,\rm jet} < $~400~GeV and rapidity $|\eta_{\rm jet}|
    <$ 
    0.7.  The plot is taken from ref.~\cite{cdf-bjets}.}
  \label{fig:cdf-bjets}
\end{figure}

An example is the inclusive $b$-jet spectrum measured by
CDF~\cite{cdf-bjets}.  Fig.~\ref{fig:cdf-bjets} shows the ratio of the
experimental measurement to the next-to-leading order (NLO) calculation
of~\cite{Frixione:1996nh}. A striking feature of this plot is the size
of the theoretical (scale variation) uncertainties ($\sim 50\%$).
One notes in particular that there is a significant region where the
experimental uncertainties are smaller than the theoretical ones.
Furthermore, the $b$-jet theory uncertainties are considerably larger
than the corresponding ones for the normal (light) jet inclusive
spectrum ($\sim 10$-$20\%$), see for example~\cite{cdf-jets}.

\begin{figure}[tp]
  \centering
  \includegraphics[height=1.0\textwidth,angle=270]{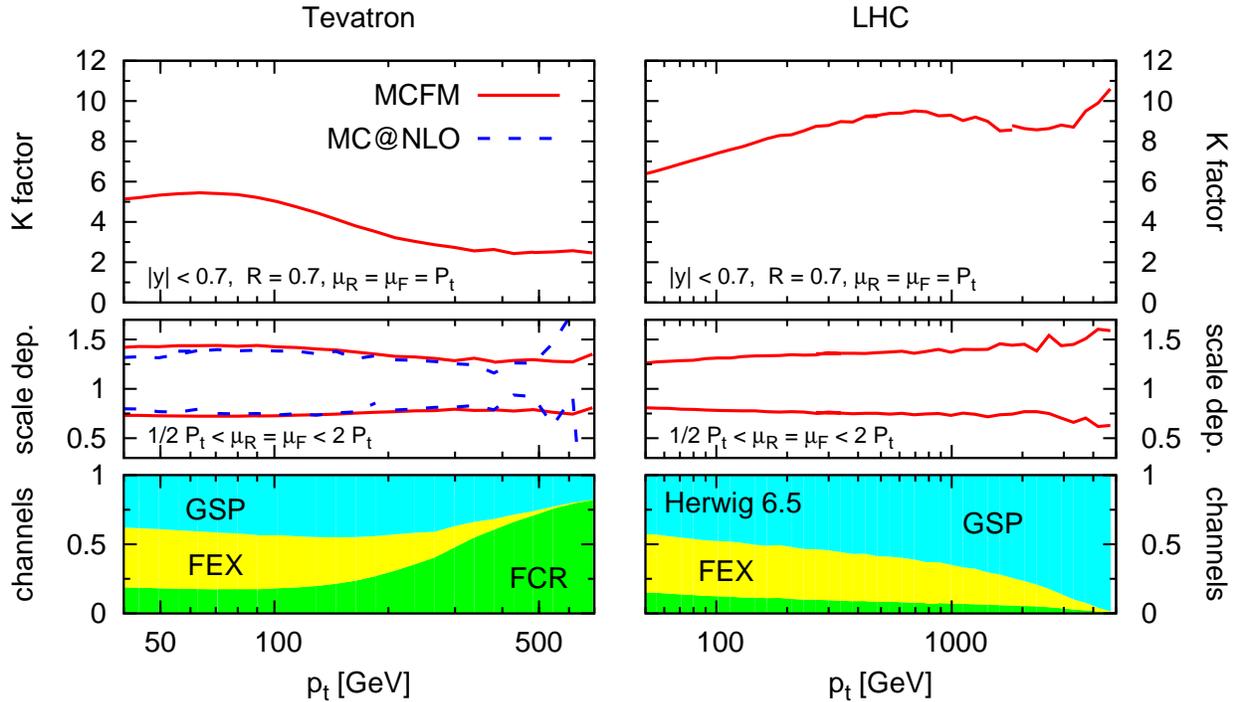}
  \caption{Top: $K$-factor for inclusive $b$-jet spectrum as computed with 
    MCFM~\cite{mcfm}, clustering particles into jets using the $k_t$
    jet-algorithm~\cite{kthh}  with $R$=0.7, and selecting jets in the central
    rapidity region ($|y| <0.7$). Middle: scale dependence obtained by
    simultaneously varying the renormalisation and factorisation
    scales by a factor two around $\pt$, the transverse momentum of
    the hardest jet in the event. Bottom: breakdown of the Herwig
    \cite{Herwig} inclusive
    $b$-jet spectrum into the three major hard underlying channels
    cross sections (for simplicity the small $bb \to bb$ is not
    shown).}
  \label{fig:kfact+channels}
\end{figure}

The origin of the large theoretical uncertainties in
fig.~\ref{fig:cdf-bjets} can be understood by examining
fig.~\ref{fig:kfact+channels}. Its top panels show the $K$-factor
(NLO/LO) as obtained with MCFM for the Tevatron Run II ($p\bar p$,
$\sqrt{s}= 1.96$ TeV, left) and for the LHC ($pp$, $\sqrt{s}= 14$ TeV,
right).\footnote{Fig.~\ref{fig:cdf-bjets} has been obtained using a
  midpoint type~\cite{Blazey:2000qt} cone algorithm, however given the
  recent discoveries~\cite{TeV4LHC,SISCone} of infrared safety issues
  in midpoint cone algorithms, we prefer to illustrate our arguments
  with an inclusive $k_t$-algorithm~\cite{kthh}. In practice, we
  expect most features of the figure to be insensitive to the choice
  of algorithm, for example also with an infrared safe cone-type algorithm
  such as SISCone~\cite{SISCone}.}
The fact that the $K$-factor is considerably larger than one indicates
that the perturbative series is very poorly convergent, and implies
that the NLO result cannot be an accurate approximation to the full
result.
It is for this reason that the scale dependence (middle panels) is
large. One might think that a calculation with
MC@NLO~\cite{Frixione:2003ei} should do better, since it includes both
NLO and all-order resummed logarithmically enhanced terms. This turns
out not to be the case, as can be seen from its persistently large
scale dependence.\footnote{Poor numerical convergence prevented us
  from presenting the scale dependence for MC@NLO at the LHC. Note
  also that no $K$-factor has been shown for MC@NLO because the LO
  result is not unambiguously defined.} %
Essentially, while MC@NLO contains a good matching between the NLO
$b$-production calculation and the $b$-quark fragmentation logarithms
in Herwig, it does not match with the logarithmic enhancements
contained in Herwig for $b$-quark production, but rather just replaces
them with the NLO result.

The poor convergence of the perturbative series is related to the
different channels for heavy quark production. At leading order (LO),
only the so-called flavour creation channel (FCR) is present, $\ell
\ell \to b\bb$, where $\ell$ is a generic light parton (quark or
gluon).
At NLO, two new channels open up, often
referred to as the flavour excitation (FEX) and gluon splitting
channels (GSP).\footnote{It is sometimes stated that it makes no
  sense, beyond LO, to separately discuss the different channels, for
  example because diagrams for separate channels interfere. However,
  each channel is associated with a different structure of logarithmic
  enhancements, $\ln^n(p_t/m_b)$, and so there is distinct physical
  meaning associated with each channel. Furthermore one can give a
  precise, measurable definition to each channel, \eg using an
  exclusive variant of the
  flavour jet algorithm discussed below. Though there will be some
  arbitrariness in any such definition, relating to the choice of
  parameters of the jet algorithm, this arbitrariness is no more
  troubling conceptually or practically than the jet-definition
  dependence that arises when determining the number of jets in an
  event.} %
In the former, a gluon from one of the incoming
hadrons splits collinearly into a $b\bb$-pair and one of those
$b$-quarks enters the hard $b\ell \to b\ell$ scattering. In the gluon
splitting process, the hard scattering process is of the form $\ell
\ell \to \ell \ell$, and one of the final-state light partons (at NLO
always a gluon) splits collinearly into a $b\bb$-pair (a jet
containing both $b$ and $\bar b$ is considered to be a $b$-jet in
standard definitions).
The various channels can be conveniently separated with a parton
shower Monte Carlo generator such as Herwig~\cite{Herwig}, where one
can determine the underlying hard channel from the hard process in the
Herwig event record.\footnote{%
  The use of Herwig to label the flavour channel does not correspond
  to a directly measurable definition, however Herwig does have the
  correct logarithmic enhancements for each channel, and furthermore
  gives results rather similar to those based on a flavour-channel
  classification using the algorithm of section~\ref{sec:algo} with
  $R=1$ (the value of $R$ that places initial and final-state
  radiation on the same footing for $k_t$ type jet algorithms).%
} %
Their relative contributions to the total $b$-jet
spectrum are shown in the bottom panel of
fig.~\ref{fig:kfact+channels}.  One sees that the supposedly LO channel
(FCR) is nearly always smaller than the two channels that at fixed
order enter only at NLO (FEX and GSP). This is because both NLO
channels receive a strong enhancement from collinear logarithms, going
as $\as^2 (\as \ln(\pt/\mb))^n$ for flavour excitation~\cite{DGLAP}
and $\as^2 \cdot \as^n \ln^{2n-1}(\pt/\mb)$ for gluon splitting ($n\ge
1$)~\cite{bmult}.

Three approaches come to mind for increasing the accuracy of the
$b$-jet spectrum prediction. The most obvious (and hardest) is to
carry out the full massive next-to-next-to-leading order calculation.
Aside from being beyond the limit of today's technology, such an
approach would still leave many of the higher order logarithms
uncalculated and so would only partially improve the situation. A
second approach would be to carry out the explicit resummation of both
the incoming and outgoing collinear logarithms. The technology for
each resummation on its own is well-known at next-to-leading
logarithmic accuracy (NLLA)~\cite{DGLAP,bmult}, though significant
effort would probably be necessary to assemble them together
effectively.
In both of the above approaches, the largest residual uncertainties
are likely to be associated with the channel with the most logarithms,
gluon splitting. This channel however does not even correspond to
one's physical idea of a $b$-jet, \ie one induced by a hard $b$-quark
and it seems somehow unnatural to include it at all as part of one's
$b$-jet spectrum.

We therefore propose a third approach to improving the accuracy of the
prediction of the $b$-jet spectrum. It is a two-pronged approach.
Firstly, one uses a definition of $b$-jets which maintains the
correspondence between partonic flavour and jet flavour. Specifically,
we take the flavour-$k_t$ algorithm of~\cite{jetflav}. Within this
algorithm, described in section~\ref{sec:algo}, a jet containing equal
number of $b$ quarks and $b$ antiquarks is considered to be a light
jet, so that jets that contain a $b$ and $\bb$ from the gluon
splitting channel do not contribute to the $b$-jet spectrum. The use
of this kind of algorithm already leads to some reduction of the
theoretical uncertainty on the $b$-jet spectrum with a standard
massive calculation (\eg with MCFM).
Further improvement can be obtained by exploiting the fact that the
logarithms of $\pt/\mb$ that remain are those associated with flavour
excitation, which coincide with those  resummed in the
$b$-quark parton distribution function (PDF) at scale $\pt$. If one
uses a $b$-quark PDF to resum these logarithms, no other logarithms
$\ln(\pt/\mb)$ appear in the rest of the calculation, so that one can
safely take the limit $\mb\to0$ and one misses only corrections
suppressed by powers of $(\mb/\pt)^2$ (possibly with additional
logarithms).
The validity of this procedure is a consequence of the infrared safety
of the jet-flavour even in the massless limit (see later).\footnote{We
  note that such a `$5$-flavour scheme', with resummed $b$-quark PDFs
  has been used before in MCFM for $H+b$ and $Z+b$
  production~\cite{MCFMbZ}. In that case, because a non-flavour jet
  algorithm was used, it was necessary to supplement the results with
  an explicit massive calculation of the NLO gluon-splitting process.
  We thank John Campbell for bringing this to our attention.}
This third approach is therefore the one which is
technically the easiest to pursue and which should simultaneously
reduce the theoretical uncertainties the most. In
section~\ref{sec:res} we present results for $c$-and $b$-jets using
this method.
A similar approach can be used also in different contexts, \eg
recently the flavour-algorithm of~\cite{jetflav} has been used to
define the $\ee$ forward-backward asymmetry for $b$ in an
infrared-safe way, making it possible to compute this quantity at NNLO
using a massless QCD calculation~\cite{Weinzierl:2006yt}.

Several issues deserve detailed discussion in the above approach. Firstly for moderate values
of $\pt$ (or of the jet energy in $\ee$), finite-mass effects may not be
completely negligible. It is therefore important to determine their
size. We explain briefly how this can be done in
section~\ref{sec:res}, with further details given in appendix~\ref{sec:finite-mass-effects}. A
second issue is an experimental one related to the limited efficiency
for the identification of $B$ and $D$ hadrons. Though not strictly
within the remit of a theoretical paper, we do find it useful to
discuss various points related to this issue in
section~\ref{sec:expissues}. 
Finally we also comment on the question of electroweak effects, in
appendix~\ref{sec:ew}.

\section{The heavy-quark jet algorithm}
\label{sec:algo}
In general, flavour-algorithms provide an IR-safe definition of the
flavour of a jet, provided one knows the (light or heavy) flavour of
each parton involved. However to study heavy-quark jets it is not
necessary to know the flavour of light quarks, because gluons and
light flavoured quarks can be considered as flavourless, while one
assigns to heavy (anti)-quarks a flavour 1 (-1). We define the
heavy-flavour of a (pseudo)-particle or a jet as its net heavy flavour
content, \ie the total number of heavy quarks minus heavy antiquarks.
One may alternatively use the sum of the number of quarks and
anti-quarks modulo 2.  Flavourless (flavoured) objects are then those
with (non-)zero net flavour.
We present here the inclusive version of the heavy-flavour jet
algorithm for hadron-hadron collisions, referring the reader
to~\cite{jetflav} for the motivation of the formalism (as well as the
original exclusive formulation):

\begin{enumerate}
\item For any pair of final-state particles $i$, $j$ define a class of
  longitudinal boost invariant distances $d_{ij}^{(F,\alpha)}$
  parametrised by $0<\alpha\le 2$ and a jet radius $R$
  \begin{equation}
    \label{eq:dij-flavour-alpha}
    d_{ij}^{(F,\alpha)} = \frac{\Delta y_{ij}^2 + \Delta \phi_{ij}^2}{R^2}
    \times\left\{ 
      \begin{array}[c]{ll}
        \max(k_{ti}, k_{tj})^\alpha \min(k_{ti}, k_{tj})^{2-\alpha}\,,
        & \> \mbox{softer of $i,j$ is flavoured,}\\
        \min(k_{ti}^2, k_{tj}^2)\,, & \> 
        \mbox{softer of $i,j$ is flavourless,}
      \end{array}
    \right.
  \end{equation}
  where $\Delta y_{ij} = y_i - y_j$, $\Delta\phi_{ij} = \phi_i -
  \phi_j$ and $k_{ti}$, $ y_i$ and $\phi_i$ are respectively the
  transverse momentum, rapidity and azimuth of particle $i$, with
  respect to the beam.
  
  For each particle define a distance with respect to the beam $B$ at
  positive rapidity,
  \begin{equation}
    \label{eq:diB-flavour-alpha}
    d_{iB}^{(F,\alpha)} = \left\{
      \begin{array}[c]{ll}
        \max(k_{ti}, k_{tB}( y_i))^\alpha 
        \min(k_{ti}, k_{tB}( y_i))^{2-\alpha}
        \,, & \quad\mbox{$i$ is flavoured,}\\
        \min(k_{ti}^2, k_{tB}^2( y_i))\,, & \quad\mbox{$i$ is flavourless,}
      \end{array}
    \right.
  \end{equation}
with 
\begin{equation}
  \label{eq:ktB}
  k_{tB} ( y) =
  \sum_i k_{ti} \left( \Theta( y_i -  y) +
    \Theta( y -  y_i) e^{ y_i -  y}\right)\,.
\end{equation} 
Similarly define a distance to the beam $\bar B$ at negative rapidity
by replacing $k_{tB}$ in eq.~\eqref{eq:diB-flavour-alpha} with $k_{t
  \bar B}$
\begin{equation}
  \label{eq:ktBbar}
  k_{t\bar B} ( y) = 
  \sum_i k_{ti} \left( \Theta( y -  y_i) +
    \Theta( y_i -  y) e^{ y -  y_i}\right)\,.
\end{equation}

\item Identify the smallest of the distance measures. If it is a
  $d_{ij}^{(F,\alpha)}$, recombine $i$ and $j$ into a new particle,
  summing their flavours and 4-momenta; if it is a
  $d_{iB}^{(F,\alpha)}$ (or $d_{i\bar B}^{(F,\alpha)}$) declare $i$ to be
  a jet and remove it from the list of particles.
 \item Repeat the procedure until no particles are left.
\end{enumerate}
Sensible values for $\alpha$ are $1$ or $2$ \cite{jetflav} and $R$
should both be kept of order $1$, to avoid the appearance of large
logarithms of $R$.

The IR-safety of this algorithm was proved in~\cite{jetflav}. A
general consequence of IR-safety is that it allows one to take the
limit $m_Q^2\to 0$ (any finite-mass corrections being suppressed by
powers of $m_Q^2/\pt^2$) as long as collinear singularities
associated with incoming heavy quarks are factorised into a heavy
quark PDF.
This means that we can compute heavy-quark jet cross sections using a
simpler, light-flavour NLO program, rather than a heavy-flavour
one~\cite{Mangano:1991jk}. Furthermore IR and collinear safety ensure
that one obtains the same results whether one considers heavy-quark
flavour at parton level, or heavy-meson flavour at hadron level,
modulo corrections suppressed by powers of $\Lambda_{QCD}/\pt$.

\section{Results}
\label{sec:res}

\begin{figure}[tp]
  \centering
  \includegraphics[width=0.59\textwidth,angle=270]{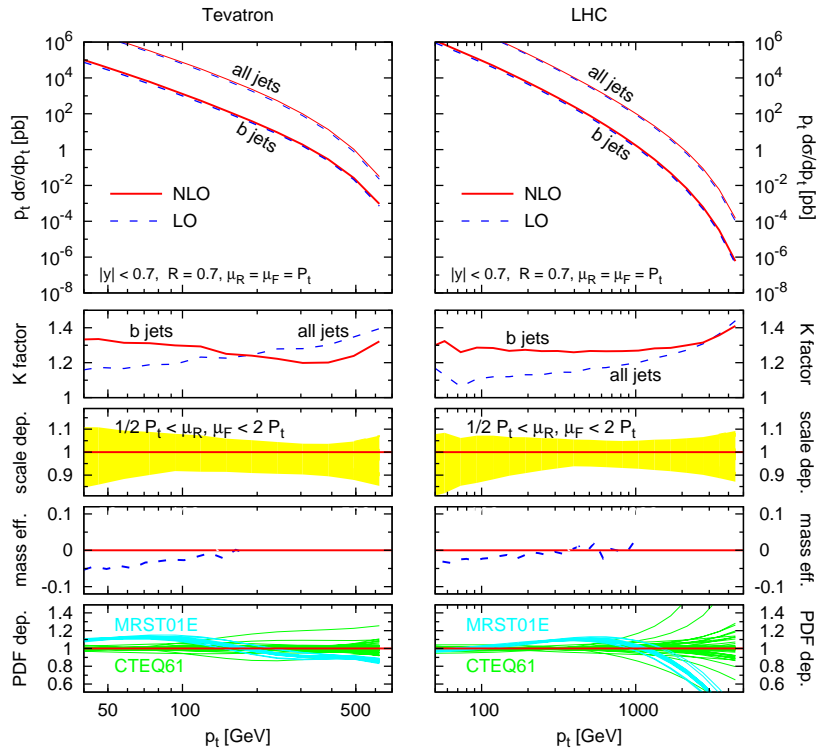}
  \caption{Inclusive jet spectrum at the Tevatron (right) and at the
    LHC (left). The top two panels show results for both $b$-jets and
    all-jets, while the lower three panels apply only to $b$-jets.
    See text for further details.}
  \label{fig:two-spect}
\end{figure}

In Fig.~\ref{fig:two-spect} we present the inclusive $b$-jet
$\pt$-spectrum as obtained with the flavour algorithm specified above.
We have used the jet-algorithm parameters $\alpha=1$, and $R=0.7$, the
latter having been shown to limit corrections associated with the
non-perturbative underlying event~\cite{cdf-jets}. The left (right)
column of the figure shows results for the Tevatron run II (LHC). We
have selected only those jets with rapidity $|y| <0.7$. We also show
the full inclusive jet spectrum (all jets) as obtained with a standard
inclusive $k_t$-algorithm with $R=0.7$.

The spectra have been calculated using NLOJET~\cite{NLOJET}. The
publicly available version sums over the flavour of outgoing partons.
We therefore had to extend it so as to have access to the flavour of
both incoming and outgoing partons.
We fixed the default renormalisation and the factorisation scales to
be $\Pt$, the transverse momentum of the hardest jet in the event and
chose as a default PDF set CTEQ61m~\cite{Stump:2003yu}.
We also used the a posteriori PDF library (APPL)
of~\cite{Carli:2005ji}, together with the HOPPET~\cite{HOPPET} and
LHAPDF~\cite{LHAPDF} packages to allow us to vary scales and PDF sets
after the NLOJET Monte Carlo integration.

The figure shows the inclusive jet spectrum at LO (blue, dashed) and
at NLO (red, solid) for all jets and for $b$-jets. The $b$-jet cross
section is always a few percent of the light jet one.  The $K$-factor,
the ratio of NLO over LO cross-section is shown below and is similar
(between 1.15 and 1.4) for light and $b$-jets, both at the Tevatron
and at the LHC.
To provide an estimate for the theoretical uncertainty we vary separately
the factorisation and the renormalisation scale in the range $1/2 \Pt
< \mu_R, \mu_F < 2 \Pt$. The band associated with this variation is
shown in the plots below. We see that this is at most a 15\% effect in
the region considered. We note that our procedure is more conservative than the
usual simultaneous variation of $\mu_R$ and $\mu_F$ (as done in
figures~\ref{fig:cdf-bjets} and \ref{fig:kfact+channels}).

We have also calculated (but do not show) the $b$-jet spectrum for our
definition of heavy jets using a massive NLO calculation with
MCFM~\cite{mcfm}.  We find that the results are consistent with those
from the massless calculation, though the uncertainties in the massive
calculation are much larger, only slightly smaller than those in
fig.~\ref{fig:kfact+channels}.

\begin{figure}[t]
  \centering
  \includegraphics[width=0.6\textwidth,angle=270]{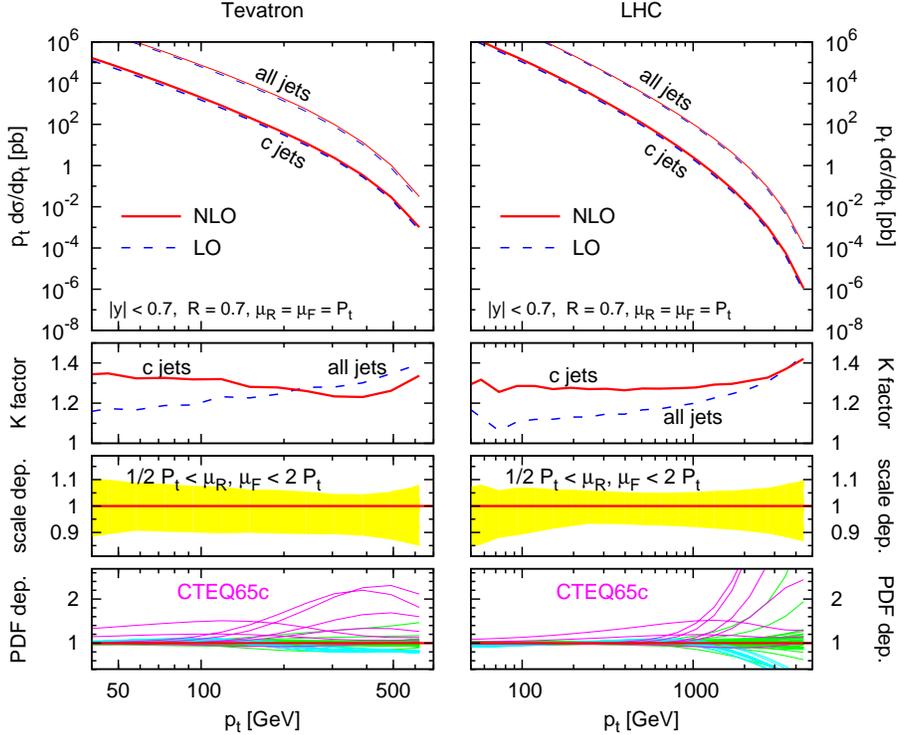}
  \caption{Inclusive jet spectrum at the Tevatron (right) and at the LHC 
    (left) for generic jet production and for $c$-jet production. See
    text for further details.  }
  \label{fig:two-spect-c}
\end{figure}

Though the massive calculation is not itself of much direct interest
given its significant uncertainties, it does enable one to
estimate residual finite-mass effects, via the relation
\begin{equation}
  \label{eq:massless-to-massive}
  \left. \frac{d\sigma_b}{dp_t} \right|_{m=m_b}
   = 
   \frac{d\sigma_b^\mathrm{nlojet}}{dp_t}
    + \lim_{m_0\to 0} \left(
      \left. \frac{d\sigma_b^\mathrm{MCFM}}{dp_t} \right|_{m=m_b}
      - \left. \frac{d\sigma_b^\mathrm{MCFM}}{dp_t} \right|_{m=m_0}
      + C(p_t) \ln \frac{m_b}{m_0}
    \right).
\end{equation}
Here, the contents of the bracket corresponds to the evaluation of the
difference between the result for the true mass and the massless
limit, while subtracting logarithms such that the massless MCFM
calculation is effectively being carried out with a coupling and
$b$-PDF that are mass-independent at scale $p_t$. 
Further details and the form for $C(\pt)$ are provided
in appendix~\ref{sec:finite-mass-effects}. The relative size of the
residual finite mass effects 
is shown in the penultimate panel of fig.~\ref{fig:two-spect}. They
decay somewhat more slowly with $p_t$ than the naive expectation of
$m_b^2/p_t^2$ (a feature noted before in \cite{Cacciari:1998it}), perhaps
because they have logarithmic enhancements.  Nevertheless they are
always below $6\%$ and given their modest size compared to the massless
perturbative uncertainties, we choose not to explicitly add them to
the main NLOJET results.

To illustrate the dependence on the parton densities we show in the
bottom panel of fig.~\ref{fig:two-spect} the effect of using all
members of the CTEQ61~\cite{Stump:2003yu} and
MRST2001E~\cite{Martin:2002aw} PDF sets, relative to the default
CTEQ61m choice.\footnote{We have also examined the
  CTEQ65 \cite{Tung:2006tb}, MRST2004nlo and
  MRST2004nnlo \cite{Martin:2004ir} sets and found similar results.}
We see that the effect is always moderate at the Tevatron ($\lesssim
20\%$), while it is large at the LHC in the high $\pt$ region,
presumably because the $b$ and gluon PDFs are not well constrained in
that region.

We have also calculated the spectrum for charm jets and the results
are shown in figure~\ref{fig:two-spect-c}. We omit the panel showing
finite-mass effects because of the low charm quark mass. The most
notable difference relates to the PDF dependence. There has been some
discussion of a possible intrinsic charm (IC) component of the proton
and a recent analysis provides PDF sets, CTEQ65c, with various models
for such a component, see \cite{Pumplin:2007wg} and references
therein.
One sees that at moderate $\pt$ these sets suggest that there is up to
$40\%$ uncertainty in the charm jet spectrum and at higher $\pt$ the
uncertainty reaches a factor two.  Further investigation reveals that
the moderate $\pt$ uncertainty is related to a possible sea-like IC
component. In the sea-like scenario considered
in~\cite{Pumplin:2007wg} it was assumed that charm and anti-charm are
distributed as the up and down sea components in the proton. At higher
$\pt$ the uncertainty is due to the valence-type models for IC
considered in~\cite{Pumplin:2007wg}, specifically the original BHPS
light-cone model~\cite{Brodsky:1980pb}, and a meson-cloud
picture~\cite{Navarra:1995rq} in which the IC arises from virtual
low-mass meson+baryon components of the proton.

\begin{figure}[t]
\centering
  \includegraphics[width=0.63\textwidth,angle=270]{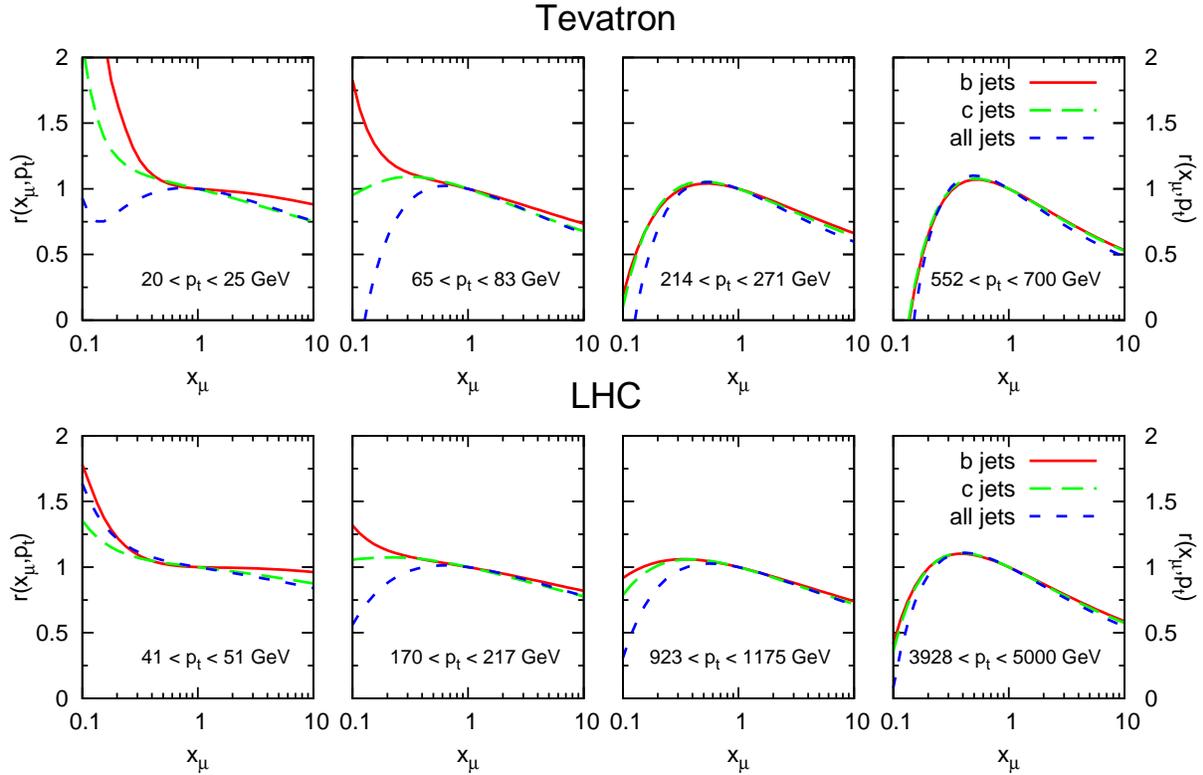}
\caption{Ratio of spectrum at factorisation and renormalisation 
    scale $\mu_R=\mu_F= x_\mu \Pt$ and at $\mu_R=\mu_F=\Pt$.}
  \label{fig:two-muline}
\end{figure}

Let us now return to the question of theoretical uncertainties in our
predictions, specifically the scale dependence.
Fig.~\ref{fig:two-muline} shows the ratio
\begin{equation}
  \label{eq:r}
r(x_\mu, \pt) =\frac{\frac{d\sigma}{d\pt}(\mu=x_\mu\, \Pt)}
{\frac{d\sigma}{d\pt}(\mu=\Pt)}\,,
\end{equation}
for the inclusive and heavy-quark jet cross-sections in various
$\pt$-bins. The factorisation and renormalisation scales are
varied simultaneously, $\mu_R=\mu_F= \mu=x_\mu \Pt$.
At low $\pt$ at the Tevatron and at intermediate $\pt$ at the LHC the
scale dependences are quite different at low values of $x_\mu$ ($\lesssim
0.5$) due to the dominance of different partonic channels.  However,
the sensitivity, \ie the dependence of $r(x_\mu, \pt)$ on $x_\mu$ remains
always of the same order for heavy-quark jets and all jets. The charm
ratio is generally intermediate between the $b$ and all-jet ratios, as
is natural given the relative masses of the charm and bottom quarks.

The fact that the scale dependences are similar for all and heavy jets
in many of the $\pt$ bins, suggests that if one considers the ratio of
heavy to all jets a significant part of the theory uncertainties may
cancel. Additionally, a number of experimental uncertainties may
cancel, for example part of the jet energy scale and luminosity
dependence. 

\begin{figure}[p]
  \centering
  \includegraphics[width=0.55\textwidth,angle=270]{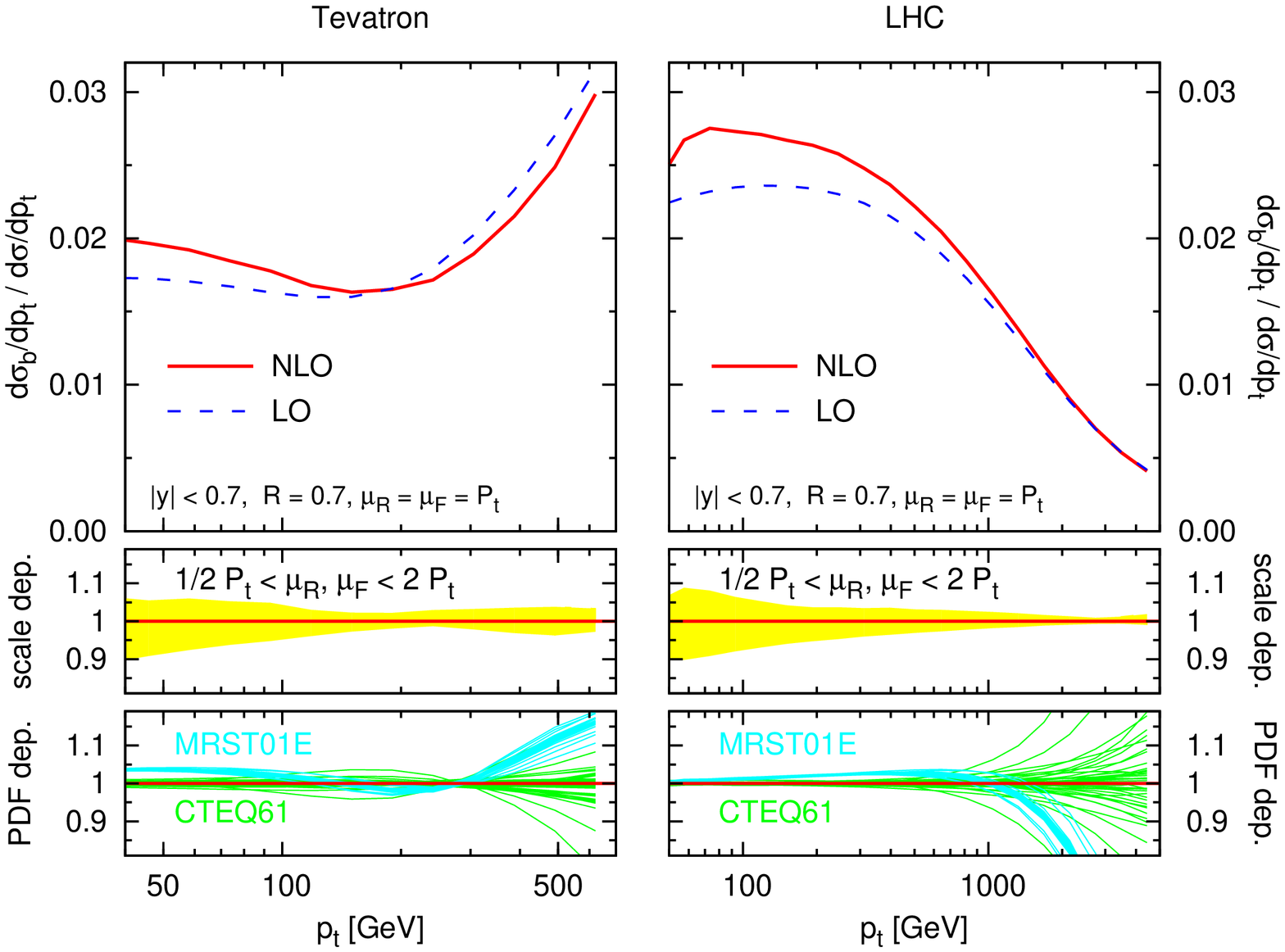}\hfill
  \\[15pt] 
  \includegraphics[width=0.55\textwidth,angle=270]{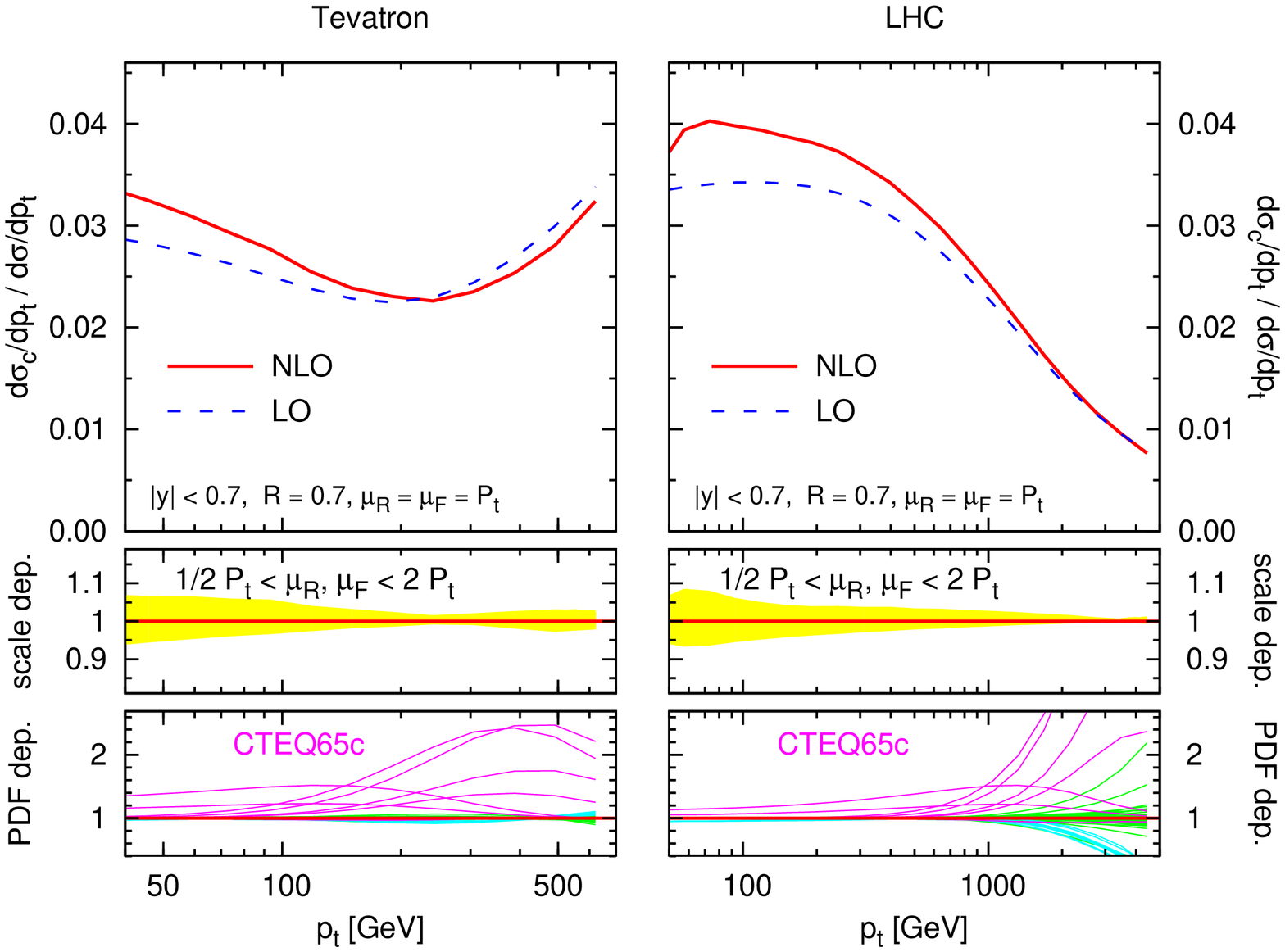}
  \caption{Top: ratio of $b$-jet to inclusive jet spectra at the
    Tevatron and at the LHC. Bottom: ratio of the $c$-jet to inclusive
    jet spectra. Further details are provided in the text.}
  \label{fig:two-ratios}
\end{figure}

Accordingly in fig.~\ref{fig:two-ratios} we show the ratio of $b$- and
$c$-jet spectra to the all-jet spectra. The ratio is always of the
order of a few percent and is somewhat larger for $c$-jets than for
$b$-jets, as is to be expected given the larger charm PDF. At higher
$\pt$ it increases at the Tevatron and decreases at the LHC due to the
different behaviour of the PDFs in the range of $x$ and $Q^2$ probed
by the two different machines. 
In particular, at large $x$ all-jet spectra are dominated by channels
with valence incoming quarks. The same is true at the Tevatron for
heavy-quark jets, where the main high-$p_t$ production channel is
$q\bar q \to Q \bar Q$.
At the LHC, on the contrary, high-$p_t$ heavy quarks are produced
mainly via $Qq \to Qq$ processes, so that heavy-jet spectra are
suppressed by the heavy-quark PDF.

The lower panels of fig.~\ref{fig:two-ratios} show the uncertainty
associated with the variation of factorisation and renormalisation
scales and the PDFs.  The scale dependence is reduced in the whole
$\pt$ range compared to that for the heavy-jet spectra. This is
especially the case at large $\pt$ (cf. fig.~\ref{fig:two-muline}).
The PDF dependence is also reduced except in the case of charm jets
using PDFs with an intrinsic charm component, CTEQ65c.

\section{Experimental issues}
\label{sec:expissues}
The main outstanding question is that of the experimental
measurability of heavy flavour jets as defined here. We examine this
specifically for $b$-jets, since they have been much more widely
studied. We will comment briefly on $c$-jets at the end of the
section.

The question of the experimental measurability of $b$-jet definition
can only truly be settled by a detailed experimental study. However
several points lead us to believe that such a measurement might well be
possible.
Our discussion here is inspired in part by that
in~\cite{Acosta:2004nj}, which measured $B\bar B$ azimuthal
correlations at the Tevatron, including the region of small angular
separation between the $B$ and $\bar B$, which is the experimentally
non-trivial region also for our definition of $b$-jets.  One should be
aware in the discussion below that the
correspondence between our needs and what was done
in~\cite{Acosta:2004nj} is only partial, insofar as the measured
$B$-hadrons were not used as inputs to a jet algorithm, and also had
lower typical transverse momenta than would the $B$-hadrons in $b$-jet
studies.\footnote{At higher $p_t$'s the fraction of $b\bb$ pairs at
  small angles will be increased, making the analysis more difficult,
  on the other hand the secondary vertices will be displaced further
  from the primary vertex and this should facilitate the analysis. The
  extent to which these two effects cancel can only be determined by a
  full experimental study.}

In an ideal world the input to the jet flavour algorithm would be a
list of momenta of all particles in the event together with
information about which particles correspond to a $B$-hadron. We are
allowed to use $B$-hadrons rather than $b$ quarks in the algorithm
because the flavoured jets are infrared and collinear safe and the
fragmentation of a $b$ quark into a $B$-hadron should have no more
effect than collinear radiation from the $b$ quark.

Experimentally one has information on charged tracks and their
momenta, calorimeter energy deposits, and $b$-tags. The latter
typically exploit the long lifetime of $B$-hadrons, which causes the
$B$-hadron to decay some small but measurable distance away from the
primary interaction vertex of the event. If the $B$-hadron decay
products include two or more charged particles then a secondary vertex
may be identified from the intersection of the resulting charged
tracks, whereas if the decay involves only one charged track then one
may still obtain a $b$-tag based on the finite impact parameter
between that track and the primary vertex. Often the $b$-tagging is
restricted to tracks within hard jets, so as to reduce certain
backgrounds.

Current $b$-tagging abilities don't correspond to our `ideal world'
scenario for a variety of reasons.  Firstly, since one often sees only
a subset of the $B$-hadron decay products, one does not know the
$B$-hadron momentum. This should not matter since the jet algorithm
will in its first steps recombine the observed charged tracks in the
decay with the calorimeter energy deposits from the neutral particles
(other than neutrinos) in the decay.

A second problem is that whereas experiments first search for jets and
then do the $b$-tagging, we need the information on $b$-tags before
running the jet algorithm. This should not be a major obstacle: one
may first identify jets using a standard $k_t$ or cone algorithm, with
large radius parameter (so as to catch most $b$'s, as done in
\cite{Acosta:2004nj}), carry out the $b$-tagging, and then run the
flavour algorithm using that information.

The third and potentially most serious issue relates to the finite
efficiency for $b$-tagging, and notably for double $b$-tagging inside
a single jet. The efficiency for $b$-tagging is limited for various
reasons: partly because of the need to place cuts on impact parameter
to avoid backgrounds from decays of charm hadrons, which also decay a
small but measurable distance from their production vertex (such
backgrounds are partially reduced also by using the invariant mass of
the decay products); and partly because of issues related to detector
limitations.
Double $b$-tagging for a pair of
$B$-hadrons that are close in rapidity and azimuth (\ie in the same
jet) is considered particularly difficult, because of the need to be
sure that, if one sees two secondary vertices in a jet, they aren't
`sequential tags' from a single $b$, \ie the vertex from a $B$-hadron
decaying to a $D$-hadron plus other particles, followed by the vertex
from the $D$ decay. Double $b$-tagging inside a single jet is
nevertheless possible, albeit currently with limited efficiency, as has been
shown in~\cite{Acosta:2004nj}. This is important because our algorithm
relies on jets with with two $b$'s inside being identified as light
jets.

To evaluate the impact of finite efficiencies, we consider the
following simple model. We suppose the efficiency for tagging a single
$B$-hadron to be $x$, and the efficiency for tagging two $B$-hadrons
that are well-separated (\ie in separate jets) to be $x^2$. Typical
values for $x$ are $\sim 0.5$. In contrast the probability of
tagging two nearby $B$-hadrons is taken to be $y x^2$ (while the
probability for tagging neither is $(1-x)^2$), with $y\simeq
0.2$~\cite{Acosta:2004nj} a measure of the extra difficulty of tagging
two nearby $B$-hadrons.
If, in a given bin, the number of true $b$-jets is $T$ and the number of
jets containing $b\bb$ due to gluon splitting is $G$, then the
measured number of single-tagged $b$-jets will be\footnote{We ignore
  the potential effect of a flavour-mistag on the kinematics of the
  jets. This should be justified since the differences between a
  flavour $k_t$ and a normal $k_t$ algorithm are at the level of a few
  percent in the spectra, and in the absence of flavour information
  the flavour $k_t$ algorithm just behaves like a normal $k_t$
  algorithm.}
\begin{equation}
  \label{eq:t_b_tag}
  t = x T + x (2 - (1+y) x) G\,.
\end{equation}
The contamination due to single-tagged gluon splitting is found by
taking one minus the fraction of gluon-splitting jets where neither
$b$ has been tagged, or where both $b$'s have been tagged, $x (2 -
(1+y) x) G = (1 - (1-x)^2 - x^2y)G$.
The measured number of light, `gluon-splitting', jets with double $b$
tags will be
\begin{equation}
  \label{eq:g_b_tag}
  g = x^2 y\, G\,.
\end{equation}
It is straightforward to deduce $T$ from measurements of $t$ and $g$,
\begin{equation}
  \label{eq:get_T}
  T = \frac{t}{x} - \frac{2 - (1 + y)x}{x^2 y} g\,,
\end{equation}
as long as one knows the efficiencies $x$ and $y$. In practice those
efficiencies will be imperfectly known, with uncertainties $\delta x$
and $\delta y$, and the effect of the estimated efficiency, used in
eq.~(\ref{eq:get_T}), being different from the true efficiency,
eqs.~(\ref{eq:t_b_tag}), (\ref{eq:g_b_tag}), will be an error $\delta T$
on the determination of $T$,
\begin{equation}
  \label{eq:deltaT}
  \delta T^2 = \left[(2G -T) \frac{\delta x}{x}\right]^2 + 
               \left[G(2-x) \frac{\delta y}{y}\right]^2\,,
\end{equation}
where we assume the uncertainties on $x$ and $y$ to be uncorrelated.
Since $G$ and $T$ are of the same order of magnitude (\cf
fig.~\ref{fig:kfact+channels}), the uncertainty on $T$ is essentially
given by the relative uncertainties on $x$ and $y$. If these can both
be controlled to within $10\%$\footnote{The most delicate is $y$, and
  from table~III of \cite{Acosta:2004nj}, which contains a breakdown
  of sources of systematic error (including that on the relative
  efficiency for tagging two nearby $b$'s compared to two well
  separated $b$'s), it seems that $10\%$ is a reasonable value for the
  uncertainty on $y$.}  then for $G\simeq 0.75\,T$, as we have at the
Tevatron for $p_t \sim 100 \GeV$, the relative uncertainty on $T$
should be roughly $12\%$ (for $x\simeq0.5$).  For an integrated luminosity of $2
\fb^{-1}$ there are $\sim 10^5$ events in a bin of width $10\GeV$
centred at $p_t=100\GeV$, so statistical errors will be considerably
smaller than this, and they are dominated by the relative error on $g
= x^2 y G$.  Only at higher energies, when $g$ starts to be small,
will the enhancement of relative statistical errors due to the limited
tagging efficiencies start to matter.

The above discussion is of course somewhat simplistic. In reality,
single and double $b$-tagging efficiencies may vary with rapidity, 
azimuthal separation and transverse momentum, though this ought to be
possible to account for; %
one should also correct for impurities in the $b$-tag samples -- based
on the uncertainties for the azimuthal correlations given in
\cite{Acosta:2004nj}, this may be roughly equivalent to doubling the
uncertainty on $y$; and a number of other experimental uncertainties
will also contribute, such as energy scale uncertainties.
On the other hand, steady progress is being made in $b$-tagging
techniques~\cite{cms-befficiency,atlas-befficiency,Bastos:2007nt}.
One also wonders whether the knowledge that a second $b$ is present
\emph{somewhere} in the event can be used in conjunction with a loose
second $b$-tag, so as to obtain information about where the second $b$
is most likely to be (in the same jet, in another jet, or down the
beam-pipe), giving an effectively larger value for $y$ (possibly even
$>1$). This might be important particularly when statistics are
limited, \eg at high $p_t$ and also potentially when using flavour
information in new-particle searches.

Finally, as concerns $c$-jets, though they have been the subject of
far fewer investigations, we do note that double-tag samples also
exist for charmed hadrons~\cite{doublec} and that some of the studies
on $b$-tagging~\cite{cms-befficiency} also provide information on
charm flavour, suggesting that $c$-jet studies may also be possible.
As for $b$-jets, a critical issue in a good measurement of the charm
jet spectrum will be not so much that of obtaining high tagging
efficiencies, but rather of a good understanding of those efficiencies
even if they are low.

\section{Conclusions}

The key finding of this article is that if one uses a properly defined
jet-flavour algorithm and exploits its infrared safety to take the
massless limit, predictions for heavy-quark jet spectra can be made
substantially more accurate than those based on current definitions
and NLO massive calculations (\eg MNR~\cite{Mangano:1991jk},
MCFM~\cite{mcfm} or MC@NLO~\cite{Frixione:2003ei}). When quantified in
terms of scale dependence, the QCD theoretical uncertainty is reduced from
$30-50\%$ to $10-20\%$.
This is because large higher-order logarithms that first appear at NLO
in the massive calculation are either cancelled by the jet definition
itself, or else absorbed into the heavy-quark PDF in such a way as to
become part of the leading order contribution, so that the NLO term is
truly a perturbative correction.

Measurements of the heavy-flavour jet spectra as presented here would
be of interest for a range of reasons.
Heavy-flavour jet spectra measured so far do not distinguish between
`true' heavy-flavour jets and gluon jets that fragment to $Q\bar Q$.
Our definition instead provides just the true flavoured-jet component.
Thus for the first time not only is the momentum of a hard parton a
meaningful observable quantity (as \emph{defined} by the jet
algorithm), but so is its flavour.

More generally, heavy-flavour jets, in particular $b$-jets, are used
in a variety of contexts, including PDF measurements, top quark
studies, and searches for new particles. These can only benefit from a
properly defined jet flavour.  One example seen in section~3 is for the
charm PDF: current measurements leave considerable room for a
non-perturbative `intrinsic' charm component in the proton, and given
an experimental accuracy that matched the theoretical accuracy of our
charm-jet predictions, significant constraints could be placed on this
intrinsic component. Similarly, a measurement of $W$+$c$-jet
production could help constrain the strange quark PDF~\cite{Lai:2007dq}.

To calculate the heavy-flavour jet spectra shown here, we used NLOJET.
By default it sums over the flavours of outgoing partons, so we
modified it so as to be able to disentangle the flavour information.
Though not completely trivial, this was quite a bit simpler than
writing a new NLO Monte Carlo program for a massless process, and very
much simpler than writing the corresponding heavy-flavour Monte Carlo
program. One could analogously extract the flavour information from
the many other NLO Monte Carlo programs involving massless QCD
particles, thus providing heavy-flavour jet predictions in a range of
processes.  The usefulness of the flavour information is such that we
strongly encourage NLO (and NNLO) Monte Carlo authors to provide it by
default.%
\footnote{That usefulness extends beyond the framework of the
  jet-flavour type algorithm used here. For example to improve the
  prediction for the current experimental definition of $b$-jets, one
  could use the prediction given here as a starting point and
  supplement it with an NLO ($\as^3+\as^4$) calculation of the
  difference between the experimental definition and ours, which
  starts only at $\order{\as^3}$. In principle, given the recent NLO
  calculation of the $Q\bar Q+$jet cross
  section~\cite{Dittmaier:2007wz}, the technology already exists for
  such a combination.}

To supplement the massless calculation, we also investigated residual
effects associated with the finite value of the $b$-quark mass. For
jets with $\pt \gtrsim 50\GeV$ they were of the order of $5\%$,
falling off rapidly at higher $p_t$. This was the most laborious part
of our study, however given the small size of the effects we believe
that it should be safe to neglect them in future NLO calculations of
heavy-flavour jets for other processes. Only when considering NNLO
heavy-flavour jet predictions, or low values of $\pt$ at NLO, should it become
mandatory to include finite mass effects.

The main open question remains that of the experimental usability of
our jet-flavour algorithm, mainly because of its reliance on the
correct identification of situations where a jet contains both a $B$
($D$) and a $\bar B$ ($\bar D$) hadron. As discussed in
sec.~\ref{sec:expissues}, given reasonable relative uncertainties on
single and double-tag efficiencies, we believe that it ought to be
possible to make an experimental measurement with errors that are not
disproportionate compared to theory uncertainties.  For the case of
$B$ hadrons, ongoing improvements in flavour tagging techniques,
together with the use of `loose' tagging to identify the second $B$
hadron in an event where a first $B$ hadron has already been found,
might help further. We look forward therefore to future experimental
investigations of heavy-flavour jet spectra with the definition
presented here.

\section*{Acknowledgements}
We are thankful to Matteo Cacciari, Mario Campanelli, Monica
D'Onofrio, Stefano
Frixione, Michelangelo Mangano, Andrea Rizzi, Ariel Schwartzman, Sofia
Vallecorsa and Bryan Webber for fruitful discussions.
We also thank John Campbell for assistance with MCFM.
GZ would like to thank Z\"urich University for hospitality and the use
of computer facilities.
This work was supported in part by grant ANR-05-JCJC-0046-01 from the
French Agence Nationale de la Recherche.

\appendix
\section{Finite mass effects}
\label{sec:finite-mass-effects}
Given the small theoretical errors in the predictions for heavy-quark
spectra when an infrared safe algorithm and massless calculation are
used, it is important to make sure that the error due to the massless
quark approximation remains smaller than the quoted theoretical errors
even at moderate values of $\pt$. In this appendix we explain how
$\cO{\alpha_s^2}$ and $\cO{\alpha_s^3}$ finite-mass effects can be
extracted from the massive NLO calculation in MCFM.

The procedure consists in subtracting from the full, massive NLO
result the collinear logarithms which with a massless calculation are
resummed into heavy-quark PDFs, any remainder being due to finite mass
effects $\cO{m_Q^2/\pt^2}$ potentially enhanced by logarithms.
The heavy-quark production mechanisms that
can give rise to collinear logarithms are flavour excitation and gluon
splitting.  However, if an infrared safe algorithm is used the only
logarithmic enhancements that survive are those associated with flavour
excitation.

We denote generally by $\sigma(m_Q)$ any heavy-quark jet cross section
corresponding to a set of
kinematic cuts and study its dependence on the heavy-quark
mass $m_Q$ by considering $\Delta\sigma(m_Q,m_0)=\sigma(m_Q)-\sigma(m_0)$,
where $m_0$ is an arbitrary reference mass.

When three partons are produced in the final state (NLO real
contribution) the logarithmic enhanced contribution
$\Delta\sigma(m_Q,m_0)$ due to FEX is given by
  \begin{multline}
    \label{eq:fex}
    \Delta\sigma(m_Q, m_0) 
    \>\simeq \>\frac{\as}{2\pi}\> \ln\frac{m_0^2}{m_Q^2} 
    \>\times \\
    \times \int dx_1 dx_2 \left[(P_{Qg}\otimes g)(x_1) g(x_2)
      \>\hat\sigma^{(0)}_{Qg\to Qg}(x_1,x_2)
      +g(x_1) (P_{Qg}\otimes g)(x_2) \>\hat\sigma^{(0)}_{gQ\to
        Qg}(x_1,x_2)\right.+\\
    \left.(P_{Qg}\otimes g)(x_1) q(x_2)
      \>\hat\sigma^{(0)}_{Qq\to Qq}(x_1,x_2)
      +q(x_1) (P_{Qg}\otimes g)(x_2) \>\hat\sigma^{(0)}_{qQ\to
        Qq}(x_1,x_2)\right]\,,
  \end{multline}
where the contributions in the first line are due to diagrams where
the hard scattering process is $Qg \to Qg$, while terms in the
second line correspond to diagrams where the hard scattering process
is $Qq \to Qq$ and $\hat\sigma^{(0)}_{ab\to cd}(x_1,x_2)$ denotes
the Born partonic cross section for the process $ab \to cd$ as a
function of the incoming energy fractions $x_1,x_2$.
The sums over light-quark flavours (and over quarks and antiquarks)
are implicit.

In the case of NLO virtual corrections, for calculations in which the
heavy-quark flavour is decoupled both in the running coupling and the
PDFs, the only logarithmically enhanced contribution comes from the
subprocess $q \bar q \to Q\bar Q$:
\begin{equation}
  \label{eq:virt}
  \Delta \sigma(m_Q,m_0) \>\simeq\> \frac{2 \>\as T_R}{3\pi} \>
  \ln\frac{m_0^2}{m_Q^2}\> \sigma^{(0)}_{q \bar q  \to Q \bar Q}\>.
\end{equation}
In the $gg \to Q\bar Q$ subprocess, logarithmically enhanced virtual
corrections from the renormalisation group evolution of the coupling
and the gluon distribution cancel.

\begin{figure}[t]
  \centering
  \includegraphics[width=0.45\textwidth,angle=270]{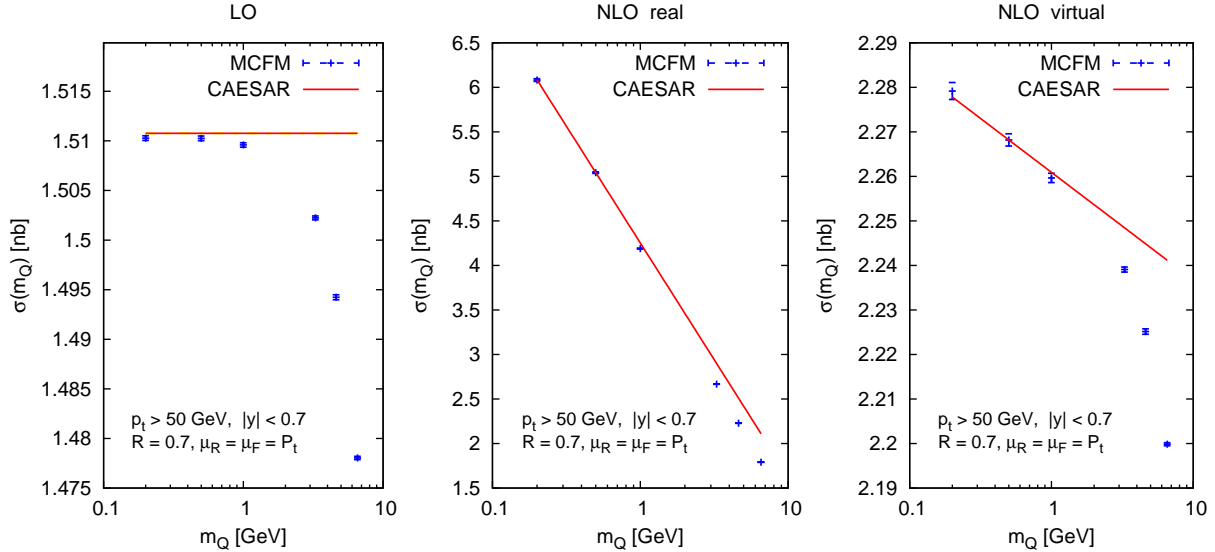}
  \caption{Various contributions to the inclusive cross section for
    $b$ jets with $p_t> 50\GeV$ and $|y|<0.7$ at the Tevatron, as a
    function of the heavy-quark mass $m_Q$. The points are from a
    massive calculation using MCFM, while at NLO the lines are given
    by the slopes in eqs.~(\ref{eq:fex},\ref{eq:virt}), with a
    constant term adjusted so as to match the massive calculation at
    $m_Q = 0.5\GeV$.}
  \label{fig:TEV-nlo-xsct}
\end{figure}

As an example, in fig.~\ref{fig:TEV-nlo-xsct} we plot $\sigma(m_Q)$,
the integrated inclusive $\pt$ spectrum at the Tevatron for $\pt > 50$
GeV and $|y|<0.7$, as a function of $m_Q$ for the real and virtual NLO
contributions, ${\cal O}(\as^3)$, as given by MCFM. We also show the
LO result for reference.

We see that in the small mass region, the cross sections computed with
MCFM are well approximated by $\sigma(m_0)+\Delta\sigma(m_Q,m_0)$,
where $\Delta\sigma(m_Q,m_0)$ are the finite-mass logarithmic
contributions in eq.~\eqref{eq:fex} and eq.~\eqref{eq:virt}, where the
integration has been performed numerically with CAESAR~\cite{caesar}.
We obtain similar results at the LHC.
Finite-mass effects for the inclusive $\pt$ spectra can then be
computed by considering the difference $\pt\, d\sigma(m_Q)/d\pt -
\pt\, d\sigma(m_0)/d\pt$, where $m_0$ is as close to zero as
numerically possible given the presence of small-mass instabilities in
the NLO calculation (we choose $m_0=0.2$ GeV at the Tevatron and
$m_0=1.0\GeV$ at the LHC) and subtracting all collinear enhancements
predicted from eqs.~\eqref{eq:fex} and~\eqref{eq:virt}.  The results
of this procedure is what is presented for $b$-jets in the 4th panel
of fig.~\ref{fig:two-spect}.  In this manner, we obtained the
coefficient to $C(p_t)$, as used in
eq.~(\ref{eq:massless-to-massive}):
\begin{equation}
  \label{eq:Cpt}
  C(p_t) \ln \frac{m_b}{m_0} = -\frac{d}{dp_t} \Delta \sigma(m_b,m_0)\,.
\end{equation}

\section{Electroweak corrections}
\label{sec:ew}

There has been discussion in the recent literature
\cite{Moretti:2006ea,Baur:2006sn,KuhnTopEW,MorettiTopEW} of
potentially large electroweak (EW) corrections to QCD light and heavy
(top) dijet cross sections.  Generally speaking there is consensus
that these effects should be modest ($\lesssim 5\%$) at the
Tevatron, but it is not uncommon for effects of up to 40\% to be
quoted at the upper end ($4\TeV$) of the $p_t$ reach of the LHC.

Two kinds of issues need to be addressed.  Firstly there are effects
that apply equally to inclusive and flavoured jet cross sections: it
has been known for some time now
\cite{Bloch-Nordsieck,Melles:2001ye,Denner:2001mn} that electroweak
loop corrections for high-$p_t$ processes involve enhancements
proportional to $\aew^n \ln^{2n}(p_t/M_W)$. Such terms are analogous to
Sudakov double logarithms in QCD, with the difference that the masses
at the electroweak scale regulate the infrared and collinear
divergences.  Because of their double logarithmic structure they
become large at high $p_t$, and they are the main culprits in the 40\%
effects quoted in~\cite{Moretti:2006ea} at the high end of the LHC
reach ($4\TeV$).  A point emphasised there is that a phenomenological
understanding of the impact of EW effects also requires that one
consider the experimental treatment of real EW radiation.
Ref.~\cite{Baur:2006sn} examined isolated $W$ and $Z$ radiation and
found that it compensated for about a quarter of the loop effects.
However, the dominant real radiation contribution should come from
(soft) collinear $W$ and $Z$ emission, and it is to be expected that
this will compensate a significant remaining part of the loop effects.

A second issue arises specifically when considering flavoured jets,
because by isolating a given flavour one breaks the electroweak SU(2)
symmetry: while the emission of a soft $W$ boson has little effect on
the energy of the jet and so should largely cancel with corresponding
virtual corrections in the inclusive jet spectrum, if the $W$ is
emitted from a $b$-jet, it will convert it into a top-quark
jet~\cite{Ciafaloni:2006qu}. %
This is often referred to as Bloch-Nordsieck
violation~\cite{Bloch-Nordsieck} and may lead to significant double
logarithmic EW corrections for the very highest $p_t$ flavoured jets
at the LHC. 
As for inclusive jet analyses, the details of the experimental
treatment are likely to be crucial, since the flavour attributed to
the jet will depend on whether the top quark is reconstructed or
whether it is only the $b$-hadron from the $t\to b+W$ decay that is
identified.
For charm jets the experimental situation will be different insofar as
the real EW emission process is $c\to s+W$, and strange hadrons do not
decay back to charmed hadrons!
The question of flavour-changing EW effects is relevant also for the
gluon splitting process, \eg $g\to c \bar c$, where one of the charm
quarks may then emit a $W$, giving a jet with a net charm flavour. If
the $W$ is not identifiable experimentally, then at high $p_t$ at LHC
this process, which has enhancements of the form $\as^m \aew^n \ln
^{2m-1+2n} (p_t/M_W)$, may give a non-negligible contribution to the
charm-jet spectrum. 

For both $b$ and $c$ jets, if the experiments prove to be able to
measure heavy flavour at these high $p_t$ values, then it will become
important to examine the above issues in more detail.

\end{document}